\documentclass{article}

\usepackage{arxiv}

\usepackage[utf8]{inputenc} 
\usepackage[T1]{fontenc}    
\usepackage{hyperref}       
\usepackage{url}            
\usepackage{booktabs}       
\usepackage{amsfonts}       
\usepackage{nicefrac}       
\usepackage{microtype}      
\usepackage{lipsum}		
\usepackage{graphicx}

\usepackage{dcolumn}
\usepackage{bm}
\usepackage{amsmath}
\usepackage{amssymb}
\usepackage{amsfonts}

\title{A generalized permutation entropy for random processes}

\date{\today}

\author{
Jos\'{e} M. Amig\'{o} \\
Centro de Investigaci\'{o}n Operativa, \\
Universidad Miguel Hern\'{a}ndez, \\
Elche, 03202 Alicante, Spain \\
\texttt{jm.amigo@umh.es}
\And
Roberto Dale \\
Centro de Investigaci\'{o}n Operativa, \\
Universidad Miguel Hern\'{a}ndez, \\
Elche, 03202 Alicante, Spain \\
\texttt{rdale@umh.es}
\And
Piergiulio Tempesta \\
Departamento de F\'{\i}sica Te\'{o}rica, \\
Facultad de Ciencias F\'{\i}sicas, \\
Universidad Complutense de Madrid, \\
28040 Madrid, Spain \\
Instituto de Ciencias Matem\'aticas, \\
28049 Madrid, Spain \\
\texttt{p.tempesta@fis.ucm.es, piergiulio.tempesta@icmat.es}
}

\begin{document}
\maketitle

\begin{abstract}
Permutation entropy measures the complexity of deterministic time series via
a data symbolic quantization consisting of rank vectors called ordinal
patterns or just permutations. The reasons for the increasing popularity of
this entropy in time series analysis include that (i) it converges to the
Kolmogorov-Sinai entropy of the underlying dynamics in the limit of ever
longer permutations, and (ii) its computation dispenses with generating and
ad hoc partitions. However, permutation entropy diverges when the number of
allowed permutations grows super-exponentially with their length, as is
usually the case when time series are output by random processes. In this
Letter we propose a generalized permutation entropy that is finite for
random processes, including discrete-time dynamical systems with
observational or dynamical noise.
\end{abstract}


\section{Introduction}
\label{sec:1}

In general, time series result from observing\ real-valued random processes
or dynamical flows at discrete times. A further step may be the
discretization of the data, a procedure called symbolic representation. Such
representations simplify the mathematical tools needed for the data analysis
and, what is more interesting for practitioners, may be sufficient for the
application sought. In this regard, ordinal patterns and permutation entropy
have become increasingly popular in nonlinear time series analysis since
their introduction by Bandt and Pompe in 2002 \cite{Bandt2002}. The reasons
are multiple. Perhaps most importantly from a theoretical point of view,
ordinal patterns, which are formally permutations, preserve the temporal
structure of a time series and, therefore, its dynamical complexity. In
fact, in one-dimensional dynamics the permutation entropy per symbol
converges to the Kolmogorov-Sinai (KS) as the pattern length grows \cite%
{Bandt2002B,Keller2010,Amigo2012}, which makes it a proxy of dynamical
entropy. From a practical point of view, the computation of permutation
entropy dispenses with ad hoc partitions, not to mention the search for
generating ones \cite{Hirata2004}. But even with real-world series, which
are usually rather noisy, tools such as permutation entropies of finite
order and the decay rate of missing ordinal patterns have proved very handy 
\cite{Unakafova2013,Unakafov2014,Carpi2010}. Further advantages of the
ordinal\ approach in the analysis of time series include speedy calculation,
robustness to noise,\ the possibility of multiscale analysis through a
varying pattern length, as well as high discriminatory power in the
classification of data, especially in combination with other complexity
indicators \cite{Rosso2007,Zunino2012,Zunino2016}. As a result, this
methodology is being successfully applied in plenty of fields, e.g., chaotic
dynamics, earth science, computational neuroscience, biomedicine,
econophysics and more; see \cite{Zanin2012,Amigo2013,Amigo2015} for recent
surveys.

More generally, the permutation entropy of a real-valued time series,
whether deterministic or random, is just the Shannon entropy of its \textit{%
ordinal representation}, i.e., the symbolic time series that results from
replacing data strings of a fixed length $L\geq 2$ by the corresponding
ordinal patterns of length $L$. There is a twist, though. Shannon's entropy
was incepted in the setting of finite-state random processes (information
sources with finite alphabets), so that the number of states (words) grows
exponentially with the length of the output (message). But in the ordinal
representation of time series, each word of length $L$ is replaced by a
permutation of $\{0,1,...,L-1\}$; if all permutations are allowed, as
happens in general with real-valued random processes (including noisy
chaotic signals), then the number of words grows super-exponentially with $L$
because $L!\simeq e^{L\ln L}$. Similarly, the number of microstates grows
super-exponentially with the number of particles in some models of
statistical mechanics, the realm of the Boltzmann-Gibbs entropy \cite%
{JPPT2018JPA}. For this \textit{super-exponential class} of processes and
many-particle systems, the Boltzmann-Gibbs-Shannon (BGS) entropy is not 
\textit{extensive}, meaning that it does not scale linearly over uniform
probability distributions or, in thermodynamical terms, at equilibrium.
Consequently, the BGS entropy per symbol or particle is unbounded and, in
general, diverges. This is the case, in particular, with the permutation
entropy for random processes.

In this Letter we propose a generalization of permutation entropy that is
finite for random processes. To this end, we resort to a new entropy
belonging to the class of group entropies, which is extensive and has
several interesting statistical properties \cite{TJ2019prep}, as well as a
normalized range. But before reaching to that point, we need to delve into 
\textit{permutation complexity}, which stands for the complexity of
discrete-time, continuous-state deterministic or random processes and their
realizations in ordinal representations \cite{Amigo2010}.

\section{Permutation complexity}
\label{sec:2}

Given a time series $(x_{t})_{t\geq 0}=x_{0},x_{1},\ldots ,x_{t},...$, with $%
t$ being discrete time and $x_{t}\in \mathbb{R}$, let $L\geq 2$ and denote
by $\mathbf{r}_{t}$ the rank vector of the string (word, block, ...) $%
x_{t}^{L}:=x_{t},x_{t+1},...,x_{t+L-1}$. That is, 
\begin{equation}
\mathbf{r}_{t}=(\rho _{0},\rho _{1},\ldots ,\rho _{L-1}),  \label{r_t}
\end{equation}%
where $\rho _{0},\rho _{1},\ldots ,\rho _{L-1}$ is the permutation of $%
0,1,\ldots ,L-1$ such that%
\begin{equation}
x_{t+\rho _{0}}<x_{t+\rho _{1}}<\ldots <x_{t+\rho _{L-1}}  \label{ord patt}
\end{equation}%
(other rules can be found in the literature). The rank vectors $\mathbf{r}%
_{t}$ are called \textit{ordinal patterns }or \textit{permutations of length}
$L$, as well as \textit{ordinal} $L$-\textit{patterns} for short; the string 
$x_{t}^{L}$ is said to be of \textit{type} $\mathbf{r}_{t}$. In case of two
or more ties in $x_{t}^{L}$, one can adopt some convention, e.g., the
earlier entry is smaller. We suppose tacitly that such occurrences are rare.
As a result, the alphabet (set of symbols) of $(\mathbf{r}_{t})_{t\geq 0}$,
the ordinal representation of the original time series $(x_{t})_{t\geq 0}$,
is the group of the $L!$ permutations of $0,1,\ldots ,L-1$, which will be
denoted by $\mathcal{S}_{L}$.

Consider a stationary, discrete-time deterministic or random process $%
\mathbf{X}=(X_{t})_{t\geq 0}$ taking values on a closed interval $I\subset 
\mathbb{R}$. By a deterministic process we mean that every output $%
(x_{t})_{t\geq 0}$ of $\mathbf{X}$ is the orbit of $x_{0}$ generated by the
same mapping $F:I\rightarrow I$, i.e., $x_{t+1}=F(x_{t})=F^{t}(x_{0})$ for $%
t\geq 0$. Therefore, random processes include deterministic ones with
observational or dynamical noise, which are instances of random dynamical
systems. Let $p(\mathbf{r})$ be the probability that a string $x_{t}^{L}$
output by $\mathbf{X}$ is of type $\mathbf{r}$ and $p=\{p(\mathbf{r}):%
\mathbf{r}\in \mathcal{S}_{L}\}$ the corresponding probability distribution.
If $p(\mathbf{r})>0$, then $\mathbf{r}$ is an \textit{allowed} \textit{%
pattern} for $\mathbf{X}$; otherwise $\mathbf{r}$ is a \textit{forbidden}
pattern. The Shannon entropy (or the BGS entropy for that matter) of $p$ is
called the (metric) \textit{permutation entropy} \textit{of order} $L$: 
\begin{equation}
H^{\ast }(X_{0}^{L})=-\sum_{\mathbf{r}\in \mathcal{S}_{L}}p(\mathbf{r})\ln p(%
\mathbf{r}),  \label{h*_mu}
\end{equation}%
where $X_{0}^{L}:=X_{0},X_{1},\ldots ,X_{L-1}$ and $0\cdot \ln 0:=0$ by
continuity. In the event that $\mathbf{X}$ is a deterministic process, $p(%
\mathbf{r})=\mu (\{x_{t}\in I:x_{t}^{L}$ is of type $\mathbf{r}\})$, where $%
\mu $ is the \textit{physical measure} of $\mathbf{X}$, which is an $F$%
-invariant measure that coincides with the empirical probability
distribution \cite{Eckmann1985}. If, otherwise, $\mathbf{X}$ is a random
process, then the probabilities $p(\mathbf{r})$ can only exceptionally be
derived from the probability distribution of $X_{t}^{L}$ \cite{Bandt2007}
so, in general, they have to be estimated, e.g., by the relative frequencies
of each $\mathbf{r}\in \mathcal{S}_{L}$ in a finite time series $%
x_{0},x_{1},....,x_{T}$,
\begin{equation}
\nu (\mathbf{r})=\frac{\#\{x_{t}^{L}\text{ of type }\mathbf{r}\in \mathcal{S}_{L}:0\leq t\leq T-L+1\}}{T-L+2},  \label{nu(r)}
\end{equation}
where $\#$ stands for \textquotedblleft number of\textquotedblright\ and $%
T\gg L!$ (maximum likelihood estimator). Then, $p(\mathbf{r}%
)=\lim_{T\rightarrow \infty }\nu (\mathbf{r})$, where this limit exists with
probability 1 when the underlying random process fulfills the following weak
condition \cite{Bandt2002}.

\textbf{Stationarity Condition}. \textit{For }$k\leq L-1$\textit{, the
probability for }$x_{t}<x_{t+k}$\textit{\ should not depend on }$t$\textit{.}

Random processes that meet this condition include, in addition to stationary
ones, non-stationary processes with stationary increments such as fractional
Brownian motion and fractional Gaussian noise. From now on we assume the
Stationarity Condition so that estimations of $p(\mathbf{r})$ converge as
the amount of data increases.

The \textit{topological permutation entropy of order} $L$ is the tight upper
bound of $H^{\ast }(X_{0}^{L})$. It is formally obtained by assuming that
all allowed $L$-patterns are equiprobable: 
\begin{equation}
H_{0}^{\ast }(X_{0}^{L})=\ln \mathcal{N}_{L}(\mathbf{X}),  \label{h*_top(X)1}
\end{equation}%
where $\mathcal{N}_{L}(\mathbf{X})$ is the number of \textit{allowed}
patterns of length $L$ for $\mathbf{X}$. In turn, the \textit{metric} and 
\textit{topological} \textit{permutation entropies} of a process $\mathbf{X}$
are obtained by taking the corresponding entropies of order $L$ per symbol
and letting $L\rightarrow \infty $,
\begin{equation}
h^{\ast }(\mathbf{X})=\lim_{L\rightarrow \infty }\frac{1}{L}H^{\ast
}(X_{0}^{L}),\;h_{0}^{\ast }(\mathbf{X})=\lim_{L\rightarrow \infty }\frac{1}{%
L}H_{0}^{\ast }(X_{0}^{L}),  \label{Rates}
\end{equation}%
provided that the limits exist. More conveniently, one can use
\textquotedblleft $\lim \sup $\textquotedblright\ (limit superior) instead
of \textquotedblleft $\lim $\textquotedblright\ to ensure that these and the
forthcoming limits converge or diverge to $+\infty $. We elaborate next on
the fact that permutation entropy is finite for deterministic processes
while diverging for random processes in general.

A mapping $F:I\rightarrow I$ is said to be piecewise monotone if there is a
finite partition of $I$ such that $F$ is continuous and monotone on each
subinterval of the partition. Let $h(F)$ be the KS entropy of 
$F$, and $h_{0}(F)$ its topological entropy \cite{Walter2000}. The following
theorem, proved in \cite{Bandt2002B}, holds.

\textbf{Theorem 1. }\textit{If }$F$\textit{\ is piecewise monotone, then (i) 
}$h^{\ast }(F)=h(F)$\textit{\ and (ii) }$h_{0}^{\ast }(F)=h_{0}(F)$\textit{.}

All one-dimensional mappings encountered in
practice are piecewise monotone, so we may assume this property for the
mappings underlying deterministic processes. Therefore, 
$h^{\ast }(\mathbf{X})\leq h_{0}^{\ast }(\mathbf{X})<\infty $ for
deterministic processes since $h_{0}(F)<\infty $ for piecewise monotone
mappings \cite{Misiurewicz1980}. Incidentally, Theorem 1(ii) implies $%
\mathcal{N}_{L}(\mathbf{X})\sim e^{h_{0}(F)L}$ ($\sim $ stands for
\textquotedblleft asymptotically\textquotedblright ), meaning that such
processes have only exponentially many allowed $L$-patterns for ever larger $%
L$'s, despite the fact that there are $L!\sim e^{L\ln L}=L^{L}$ (Stirling's
formula) possible ordinal $L$-patterns. The upshot is that the number of
forbidden patterns for deterministic processes grows super-exponentially
with $L$ \cite{Amigo2006}. Also higher dimensional dynamics along with their
lower dimensional projections may have forbidden patterns \cite{Amigo2008B}.
However, if the dynamics takes place on an attractor, so that the orbits are
dense, then the observational or dynamical noise will `destroy' all forbidden
patterns in the long run, no matter how small the noise. Theorem 1 was
generalized in \cite{Keller2019}.

On the other hand, random processes may have forbidden patterns too. For the
sake of our analysis, though, we will consider the general or `worse'
scenario in which all ordinal patterns of any length are allowed. A
necessary and sufficient condition for this is that, for $k\leq L-1$, the
probability for $x_{t}<x_{t+k}$\ is neither 0 nor 1 (so that the same holds
for $x_{t}>x_{t+k}$), which amounts to a mild addendum to the Stationarity
Condition. With this proviso, we assume hereafter $\mathcal{N}_{L}(\mathbf{X}%
)=L!$ for all random processes $\mathbf{X}$. Then, 
\begin{equation}
h_{0}^{\ast }(\mathbf{X})=\lim_{L\rightarrow \infty }\frac{1}{L}\ln
L!=\lim_{L\rightarrow \infty }\ln L=\infty  \label{h*_top(X)2}
\end{equation}%
by Stirling's formula. We conclude from (\ref{h*_top(X)2}) that permutation
entropy, unlike Shannon's entropy, cannot be applied to random processes in
general. In particular, $H^{\ast }(X_{0}^{L})$ does not scale linearly when $%
L\rightarrow \infty $ over flat probability distributions.

\begin{figure*}
\begin{center}
\includegraphics[scale=1.0]{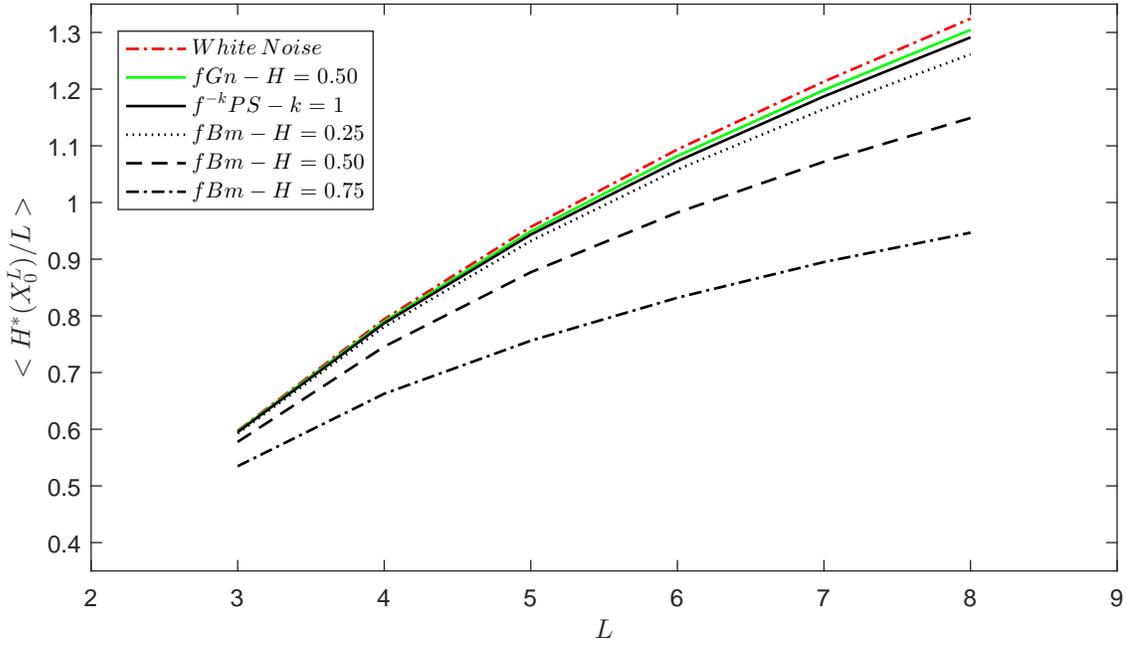}
\caption{\label{fig:Figura_1}
The average of $H^{\ast }(X_{0}^{L})/L$ over 10 realizations, $\langle H^{\ast}
(X_{0}^{L})/L\rangle $, is plotted vs $L$ for $1\leq L\leq 8$ and the
random processes listed in the inset. See the text for detail.
}
\end{center}
\end{figure*}

Numerical evidence is shown in  Fig.~\ref{fig:Figura_1}. Here we have numerically generated
10 realizations of size $T\gtrsim 50L!$ (see (\ref{nu(r)})) of the following
processes: (i) \textit{white noise} (WN) in the form of an independent and
uniformly distributed process on $[0,1]$; (ii) fractional Gaussian noise
(fGn) with Hurst exponent $H=0.5$ (Gaussian white noise); (iii) noise with
an $f^{-1}$ power spectrum (PS); (iv) fractional Brownian motion (fBm) with $%
H=0.25$ (anti-persistent process), $0.5$ (classical Brownian motion), and $%
0.75$ (persistent process). Computations were done with MatLab \cite{MATLAB2016}. The average
of $H^{\ast }(X_{0}^{L})/L$ over the 10 realizations of each process,
denoted $\langle H^{\ast }(X_{0}^{L})/L\rangle $, is then plotted against $L$%
, $3\leq L\leq 8$. We see in all cases that $\langle H^{\ast
}(X_{0}^{L})/L\rangle $ follows a seemingly divergent trajectory as $L$
grows.


\section{Group entropies}
\label{sec:3}

The theory of group entropies \cite%
{PT2011PRE,PT2016PRA,RRT2019PRA,PT2019prep} is an axiomatic approach which
allows us to construct information measures with mathematical properties
that make them suitable to describe specific universality classes of complex
systems (see \cite{JT2018ENT} for a review). We recap here some basic
definitions.

Let $\mathcal{P}_{W}$ be the set of all discrete probability distributions
with $W$ entries, i.e., $\mathcal{P}_{W}=\{p=(p_{i})_{i=1,\cdots ,W}:0\leq
p_{i}\leq 1,\,\sum_{i=1}^{W}p_{i}=1\}$. Let $S$ be a non-negative function
on $\mathcal{P}:=\cup _{W=1}^{\infty }\mathcal{P}_{W}$, so that $S$ is
defined on any probability distribution $p$ and $S(p)\geq 0$. The
Shannon-Khinchin (SK) axioms are a set of requirements first considered in 
\cite{Shannon}, \cite{Shannon2}, \cite{Khinchin} to uniquely characterize
the BGS entropy. The first three SK axioms amount to the following
properties: \newline
\noindent (SK1) $S(p)$ is continuous with respect to all variables $%
p_{1},\ldots ,p_{W}$. \newline
\noindent (SK2) $S(p)$ takes its maximum value over the uniform
distribution. \newline
\noindent (SK3) $S(p)$ is expansible: adding an event of zero probability
does not affect the value of $S(p)$.

These axioms represent a minimal set of `non-negotiable'\ requirements that
such functions $S(p)$ should satisfy necessarily to be meaningful, both from
a physical and information-theoretical point of view. Non-negative functions
on $\mathcal{P}$ that verify axioms (SK1)-(SK3) are called generalized
entropies and their structure is only known under additional conditions \cite%
{Hanel2011,Amigo2018,TJ2019prep}. Thus, the fourth SK axiom, requiring
specifically additivity on conditional distributions, leads to the BGS
entropy \cite{Khinchin},

\begin{equation}
S_{BGS}(p)=-k\sum_{i=1}^{W}p_{i}\ln p_{i},  \label{Boltzmann}
\end{equation}%
where $k$ is a positive constant that we equate to $1$ for definiteness (as
in (\ref{h*_mu})). Instead, the more general axiom of composability (see below) 
leads to the new concept of group entropy. As we will discuss shortly, this entropy, 
which includes $S_{BGS}(p)$, is better suited to deal with the diversity of themodynamical 
and complex systems. Another independent approach, based on
the concept of \textit{pseudo-additive entropy}, was formulated in \cite%
{IS2014PHYSA}.

An entropy $S(p)$ is said to be \textit{composable} if there exists a smooth
function $\Phi (x,y)$ such that%
\begin{equation}
S(p_{A}\times p_{B})=\Phi (S(p_{A}),S(p_{B}))  \label{eq:comp}
\end{equation}%
for any probability distributions $p_{A}$ and $p_{B}$, where $p_{A}\times
p_{B}$ is the product probability distribution of both. Equivalently, (\ref%
{eq:comp}) can be written as $S(A\cup B)=\Phi (S(A),S(B))$, where $A$ and $B$
are two statistically \textit{independent} subsystems of a complex system,
defined over \textit{any arbitrary probability distributions} $p_{A}$ and $%
p_{B}$, respectively, and $A\cup B$ is the system composed of $A$ and $B$.
All quantities are assumed to be dimensionless.

In addition to Equation (\ref{eq:comp}), we shall also require the following
properties for the \textit{composition law} $\Phi $: \newline
\noindent (C1) Symmetry: $\Phi (x,y)=\Phi (y,x)$. \newline
\noindent (C2) Associativity: $\Phi (x,\Phi (y,z))=\Phi (\Phi (x,y),z)$. 
\newline
\noindent (C3) Null-composability: $\Phi (x,0)=x$.

Observe that, indeed, requirements (C1)-(C3) are crucial: they impose the
independence of the composition process with respect to the order of $A$ and 
$B$, the possibility of composing three independent subsystems in an
arbitrary way, and the requirement that, when composing a system with
another one having zero entropy, the total entropy remains unchanged. In our
opinion, these properties are also fundamental: no thermodynamical or
information-theoretical applications would be easily conceivable without
these properties. For $\Phi (x,y)=x+y$ we obtain from (\ref{eq:comp}) the
additivity of the BGS entropy (\ref{Boltzmann}) with respect to the
composition of two statistically independent subsystems.

From an algebraic point of view, the requirements (C1)-(C3) define a \textit{%
formal group law} for a function (infinite series) of the form $\Phi
(x,y)=x+y\,+$ $O(2)$, where $O(n)$ stands for terms of degree $\geq n$.

\textbf{Definition 1.} \textit{A group entropy is a function }$S:\mathcal{P}%
\rightarrow \lbrack 0,\infty )$\textit{\ which satisfies the
Shannon-Khinchin axioms (SK1)-(SK3) and the composability axiom (\ref%
{eq:comp}).}

A well-known group entropy, introduced by Tsallis in \cite{TS1988JST}, is%
\begin{equation}
S_{\alpha }(p)=\frac{1}{1-\alpha }\left( \sum\limits_{i=1}^{W}p_{i}^{\alpha
}-1\right)  \label{Tsallis}
\end{equation}%
for $\alpha >0$, $\alpha \neq 1$, and $S_{1}(p):=$ $\lim_{\alpha \rightarrow
1}S_{\alpha }(p)=S_{BGS}(p)$, whose composition law is%
\begin{equation}
\Phi (x,y)=x+y+(1-\alpha )xy,  \label{PhiTsallis}
\end{equation}%
so that $S_{\alpha }(p_{A}\times p_{B})=S_{\alpha }(p_{A})+S_{\alpha
}(p_{B})+(1-\alpha )S_{\alpha }(p_{A})S_{\alpha }(p_{B})$. Except for the
Tsallis entropy, group entropies have in general non-trace forms \cite%
{ET2017JST}, that is, they cannot be written as $\sum_{i=1}^{W}g(p_{i})$,
where $g:[0,1]\rightarrow \lbrack 0,\infty )$ is a mapping with suitable
properties, usually continuity, $\cap $-convexity and $g(0)=0$ \cite%
{Hanel2011}.

As has been shown in \cite{TJ2019prep,JT2018ENT}, one can classify complex
systems according to their state space growth rate $\mathcal{W}(N)$, which counts
the number of microstates allowed as a function of the number $N$ of
particles or constituents of a given system, for large $N$. Generally
speaking, we distinguish sub-exponential, exponential and super-exponential
regimes with regard to the state space growth rate (which can be further
discriminated if necessary). All systems that are characterized by the same
asymptotic behavior of $\mathcal{W}$ define a \textit{universality class}.
According to Theorem 1 of \cite{TJ2019prep}, under mild hypotheses one can 
explicitly construct a suitable group entropy associated with a given universality class of systems, 
which would play the role of information or complexity measure for the class considered.
This specific entropy (actually, a one-parametric family of entropies) is called a $Z$%
-entropy \cite{PT2016PRA} and is denoted by $Z_{G, \alpha }(p)$,
where $G$ refers to the underlying group-theoretical structure associated with it, $\alpha >0$, 
$p\in \mathcal{P}_{W}$ and $W=\left\lfloor \mathcal{W}(N)\right\rfloor $.

Moreover, there is a \textit{unique} $Z_{G, \alpha }(p)$ entropy which is \textit{%
extensive} for the systems of a given class, that is, if $Z_{G, \alpha
}(N):=Z_{G, \alpha }(\frac{1}{W},...,\frac{1}{W})$ is the $Z$-entropy over the
uniform distribution (the most `disordered'\ situation), then 
\begin{equation}
\lim_{N\rightarrow \infty }\frac{Z_{G, \alpha }(N)}{N}=~\text{const.}
\label{extensiv}
\end{equation}%
In other words, $Z_{G, \alpha }(N)$, the topological version of $Z_{G, \alpha }(p)$, scales  
linearly with $N$, at least for $N$ sufficiently large. According
to (SK2), $Z_{G, \alpha }(p)\leq Z_{G, \alpha }(N)$ for all $p\in \mathcal{P}_{W}$.

Prototypical examples of $Z$-entropies are (i) the Tsallis entropy $%
S_{\alpha }(p)$, Equation (\ref{Tsallis}), for the sub-exponential class,
and (ii) the R\'{e}nyi entropy \cite{Renyi1961}%
\begin{equation}
R_{\alpha }(p)=\frac{1}{1-\alpha }\ln \left(
\sum\limits_{i=1}^{W}p_{i}^{\alpha }\right)  \label{Renyi}
\end{equation}%
for $\alpha >0$, $\alpha \neq 1$, and $R_{1}(p):=\lim_{\alpha \rightarrow
1}R_{\alpha }(p)=S_{BGS}(p)$, for the exponential class. Notice that $%
S_{\alpha }(p)=\frac{1}{1-\alpha }(\exp [(1-\alpha )R_{\alpha }(p)]-1)$. The 
$Z$-entropy for the super-exponential class is our next concern.

\section{A generalized permutation entropy}
\label{sec:4}

In our context, where random processes are real-valued and blocks $x_{t}^{L}$
of size $L\geq 2$ are quantized by means of ordinal $L$-patterns $\mathbf{r}%
_{t}$, discrete probability distributions $p$ refer necessarily to the
symbols $\mathbf{r}\in \mathcal{S}_{L}$ and hence the growth function is $%
\mathcal{W}(L)=L!\sim e^{L\ln L}$ under very weak conditions. This being the
case, we propose the $Z$-entropy for the super-exponential class to measure
permutation complexity. Such an entropy was introduced in \cite{JPPT2018JPA}
to describe the thermodynamic properties of the so-called pairing model,
which represents an example of a Hamiltonian system possessing a
super-exponential state space growth.

Specifically, we define the \textit{permutation} $Z$-\textit{entropy of order%
} $L$ of a process $\mathbf{X}=(X_{t})_{t\geq 0}$ as 
\begin{equation}
Z_{\alpha }^{\ast }(X_{t}^{L})\equiv Z_{\alpha }(p)=\exp \left[ \mathcal{L}%
\left( R_{\alpha }(p)\right) \right] -1  \label{Zentropy}
\end{equation}%
for $\alpha >0$, where $p\in \mathcal{P}_{L!}$ is the probability
distribution of the ordinal $L$-patterns of $X_{t}^{L}$, $R_{\alpha }(p)$ is
R\'{e}nyi's entropy (\ref{Renyi}) with $W=L!$, and $\mathcal{L}(x)$ denotes
the principal branch of the real Lambert function. $\mathcal{L}(x)$ is a
smooth function that is defined for $x\geq -1/e$ and satisfies the equation $%
\mathcal{L}(x)e^{\mathcal{L}(x)}=x$, hence $\mathcal{L}(0)=0$ and $\mathcal{L%
}(x)>0$ for $x>0$ \cite{Olver2010}. The term $-1$ in (\ref{Zentropy})
renders $Z_{\alpha }(p)=0$ in situations without uncertainty, i.e., when $%
p_{i_{0}}=1$ and $p_{i}=0$ for $i\neq i_{0}$.

From a conceptual point of view, $Z_{\alpha }(p)$ can be interpreted to be a
suitable, extensive deformation of $R_{\alpha }(p)$, sharing with it many
fundamental properties, except additivity. For example, $Z_{\alpha }^{\ast
}(X_{t}^{L})$ inherits from $R_{\alpha }(p)$ its\ $\cap $-convexity for $%
0<\alpha \leq 1$ and decreasing monotonicity with respect to $\alpha $ \cite%
{Amigo2018}, that is,%
\begin{equation}
Z_{\alpha }^{\ast }(X_{t}^{L})\geq Z_{\beta }^{\ast }(X_{t}^{L})\;\;\text{%
for\ \ }\alpha <\beta ,  \label{hierarchy}
\end{equation}%
because the function $e^{\mathcal{L}(x)}$ is strictly increasing and $\cap $%
-convex.

It is clear that $Z_{\alpha }(p)$ verifies the axioms (SK1)-(SK3) since $%
R_{\alpha }(p)$ is a group entropy and $e^{\mathcal{L}(x)}$ is strictly
increasing. The composability of $Z_{\alpha }(p)$ for the growth function $%
\mathcal{W}(L)=e^{L\ln L}$ follows from Proposition 1 of \cite{TJ2019prep}
(with $\mathcal{W}^{-1}(\xi )=\exp [\mathcal{L}(\ln \xi )]$). Alternatively,
one can directly check that if%
\begin{eqnarray}
\Phi (x,y) &=&e^{\mathcal{L[}(x+1)\ln (x+1)+(y+1)\ln (y+1)]}-1  \label{Phi}
\\
&=&x+y-\tfrac{1}{2}x^{2}-2xy-\tfrac{1}{2}y^{2}+O(3),  \notag
\end{eqnarray}%
then the composition law $Z_{\alpha }(p_{A}\times p_{B})=\Phi
(Z(p_{A}),Z(p_{B}))$ holds for any probability distributions $p_{A}$ and $%
p_{B}$.

As with conventional permutation entropy, we define the \textit{topological
permutation} $Z$-\textit{entropy of order} $L$ of a process $\mathbf{X}%
=(X_{t})_{t\geq 0}$ as the tight upper bound of $Z_{\alpha }^{\ast
}(X_{t}^{L})$, which is obtained over the uniform distribution of ordinal $L$%
-patterns: 
\begin{equation}
Z_{0}^{\ast }(X_{t}^{L})\equiv Z_{\alpha }(\tfrac{1}{L!},\ldots ,\tfrac{1}{L!%
})=\exp \left[ \mathcal{L}\left( \ln L!\right) \right] -1  \label{Ztop}
\end{equation}%
since $R_{\alpha }(\tfrac{1}{L!},\ldots ,\tfrac{1}{L!})=\ln L!$ for all $%
\alpha $. The notation $Z_{0}^{\ast }$ is justified because $\ln L!$ is
formally obtained from (\ref{Renyi}) by setting $\alpha =0$. It follows (use 
$\mathcal{L}(x\ln x)=\ln x$ for $x\geq 1/e$)
\begin{equation}
\frac{Z_{0}^{\ast }(X_{0}^{L})}{L}=\frac{e^{\mathcal{L}\left( \ln L!\right)
}-1}{L}\sim \frac{e^{\mathcal{L}\left( L\ln L\right) }-1}{L}=\frac{L-1}{L}%
\sim 1,  \label{extensiv2}
\end{equation}%
so that $Z_{\alpha }^{\ast }(p)$ is indeed extensive in the regime of
factorial growth we are interested in.

Regarding the \textit{permutation} $Z$-\textit{entropy of a random process} $%
\mathbf{X}$, 
\begin{equation}
z_{\alpha }^{\ast }(\mathbf{X})=\underset{L\rightarrow \infty }{\lim \sup }%
\frac{1}{L}Z_{\alpha }^{\ast }(X_{0}^{L}),  \label{Z*}
\end{equation}%
we highlight the following properties.

\textbf{Theorem 2}. (i) \textit{Normalized range}: $0\leq z_{\alpha }^{\ast
}(\mathbf{X})\leq 1$, where $z_{\alpha }^{\ast }(\mathbf{X})=0$ for
deterministic processes and $z_{\alpha }^{\ast }(\mathbf{X})=1$ for white
noise. (ii) \textit{Hierarchical order}: $z_{\alpha }^{\ast }(\mathbf{X}%
)\geq z_{\beta }^{\ast }(\mathbf{X})$ for $\alpha <\beta $.

To prove that $z_{\alpha }^{\ast }(\mathbf{X})=0$ for deterministic
processes, we recall that, according to Theorem 1(ii), $\mathcal{N}_{L}(%
\mathbf{X})\sim e^{h_{0}(F)L}$, where $h_{0}(F)$ is the topological entropy
of the mapping $F$ that generates $\mathbf{X}$. Therefore, if $p$ is the
probability distribution of the $L$-patterns, then $R_{0}(p)=\ln \mathcal{N}%
_{L}(\mathbf{X})\sim h_{0}(F)L$ and%
\begin{equation}
\frac{Z_{\alpha }^{\ast }(X_{0}^{L})}{L}\leq \frac{e^{\mathcal{L[}%
R_{0}(p)]}-1}{L}\sim \frac{e^{\mathcal{L}[h_{0}(F)L]}}{L}=\frac{h_{0}(F)}{%
\mathcal{L}[h_{0}(F)L]}\sim 0,  \label{proof}
\end{equation}%
where we used $e^{\mathcal{L}(x)}=x/\mathcal{L}(x)$. Furthermore, that $%
z_{\alpha }^{\ast }(\mathbf{X})\leq 1$, with equality for white noise,
follows from $Z_{\alpha }^{\ast }(X_{0}^{L})\leq Z_{0}^{\ast }(X_{0}^{L})$
and $\lim \sup_{L\rightarrow \infty }\frac{1}{L}Z_{0}^{\ast }(X_{0}^{L})=1$,
see (\ref{extensiv2}). Finally, the hierarchical order of $z_{\alpha }^{\ast
}(\mathbf{X})$ is a direct consequence of (\ref{hierarchy}).

\begin{figure*}
\begin{center}
\includegraphics[scale=0.63]{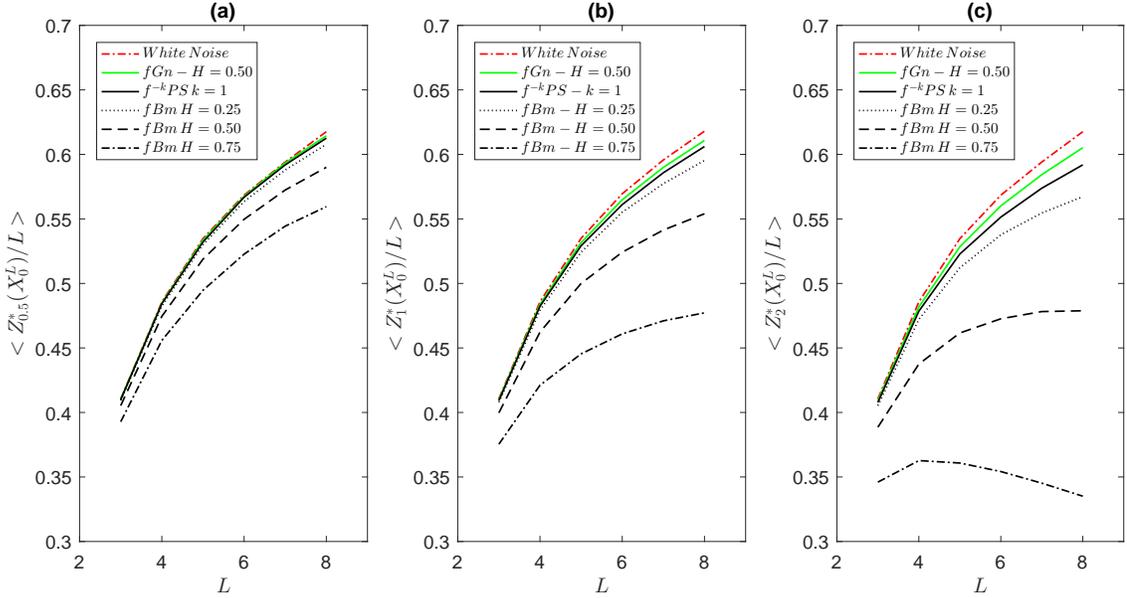}
\caption{\label{fig:Figura_2}
Same information as in Fig.~\ref{fig:Figura_1} (but notice the different scale on the Y-axis)  
for $\langle Z_{0.5}^{\ast }(X_{0}^{L})/L\rangle $ (a),
$\langle Z_{1}^{\ast }(X_{0}^{L})/L\rangle $ (b) and $\langle Z_{2}^{\ast}
(X_{0}^{L})/L\rangle $ (c). See text for detail.
}
\end{center}
\end{figure*}

As way of illustration, Fig.~\ref{fig:Figura_2} shows $\langle Z_{\alpha }^{\ast
}(X_{0}^{L})/L\rangle $, the average of the permutation entropy rate $%
Z_{\alpha }^{\ast }(X_{0}^{L})/L$ over the same 10 time series and for the
same random processes as in Fig.~\ref{fig:Figura_1}, against $L$, $1\leq L\leq 8$, where $%
\alpha =0.5$ (a), $1$ (b) and $2$ (c). Contrarily to Fig.~\ref{fig:Figura_1}, we see in all
panels of Fig.~\ref{fig:Figura_2} that $\langle Z_{\alpha }^{\ast }(X_{0}^{L})/L\rangle $
follows a seemingly convergent trajectory as $L$ grows, upper bounded by the
white noise. In agreement with (\ref{hierarchy}), $\langle Z_{0.5}^{\ast
}(X_{0}^{L})/L\rangle \geq \langle Z_{1}^{\ast }(X_{0}^{L})/L\rangle \geq
Z_{2}^{\ast }(X_{0}^{L})/L\rangle $\ for each process.

\section{Conclusions}
\label{sec:4}

Summing up, the permutation $Z$-entropy $z_{\alpha }^{\ast }(\mathbf{X})$,
Equations (\ref{Z*}) and (\ref{Zentropy}), measures the complexity of a
real-valued random process $\mathbf{X}$ through permutations, where $\mathbf{%
X}$ is supposed to fulfill the Stationarity Condition and the mild
assumption that all permutations are allowed for each length or, at least, a
super-exponentially growing number of them. First and most importantly, $%
z_{\alpha }^{\ast }(\mathbf{X})$ is always finite, contrarily to what
happens with $h^{\ast }(\mathbf{X})$, see Equation (\ref{h*_top(X)2}).
Therefore, we may claim that $z_{\alpha }^{\ast }(\mathbf{X})$ generalizes
the conventional permutation entropy to the realm of random processes.
Further distinctive features are uniqueness \cite{TJ2019prep} and the
properties listed in Theorem 2. Applications include the analysis of data in
general, and the characterization and classification of noisy signals in
particular. In this regard, the parameter $\alpha $ is an asset because it
enhances the discrimination capability of the ordinal approach. Since
real-world series are finite, one has to use permutation $Z$-entropies of
finite order $Z_{\alpha }^{\ast }(X_{0}^{L})$ in this case, where $L$ should
be chosen so as to avoid undersampling of the ordinal $L$-patterns. Compared
to $R_{\alpha }(p)$ and other entropies such as $S_{\alpha }(p)$, the
expense of computing $Z_{\alpha }(p)$ is virtually the same. On these
grounds, we propose $z_{\alpha }^{\ast }(\mathbf{X})$ as the right entropic
measure to describe the complexity of real-valued random processes in
ordinal representations.

\subsubsection*{Acknowledgments}

J.M.A. was financially supported by the Spanish Ministry of Economy,
Industry and Competitiveness (MINECO), grant MTM2016-74921-P (AEI/FEDER, EU). The
research of P.T. has been partly supported by the research project
PGC2018-094898-B-I00, MINECO, Spain, and by the ICMAT Severo Ochoa project
SEV-2015-0554 (MINECO). P.T. is member of the Gruppo Nazionale di Fisica
Matematica (INDAM), Italy.

\bibliographystyle{unsrt}

\bibliography{ZentropyArxiv_v1}

%
%
%
%

\end{document}